# Anisotropic layer-by-layer carbon nanotubes/boron nitride/rubber composite and its application in electromagnetic shielding †

Yanhu Zhan,[a] Emanuele Lago,[b,c] Chiara Santillo,[d] Antonio Esaú Del Río Castillo,[b] Shuai Hao,[e] Giovanna G. Buonocore,[d] Zhenming Chen,[f] Hesheng Xia,*[e] Marino Lavorgna*[d] and Francesco Bonaccorso,[b,g]

Multifunctional polymer composites with anisotropic properties are attracting interests as they fulfil the growing demand of multitasking materials. In this work, anisotropic polymer composites are fabricated by combining the layer-by-layer (LBL) filtration method with the alternative assembling of carbon nanotubes (CNTs) and hexagonal boron nitride flakes (hBN) on natural rubber latex particles (NR). The layered composites exhibit anisotropic thermal and electrical conductivities, which are tailored through the layer formulations. The best composite consists of four layers of NR modified with 8 phr (parts per Hundred Rubber) CNTs (~7.4 wt%) and four alternated layers with 12 phr hBN (~10.7 wt%). The composites exhibit an electromagnetic interference (EMI) shielding effectiveness of 22.41±0.14 dB mm$^{-1}$ at 10.3 GHz and a thermal conductivity equal to 0.25 W m$^{-1}$K$^{-1}$. Furthermore, when the layered composite is used as an electrical thermal heater the surface reaches a stable temperature of ~103 °C in approx. 2 min, with an input bias of 2.5 V.

## Introduction

Electromagnetic (EM) radiations emitted by electronic devices is certainly negative for equipment operation. [1-3] For this reason, it is of paramount importance to develop new materials with a tailored electromagnetic interference (EMI) shielding properties, able to contrast the electromagnetic pollution in electromagnetic sensitive applications, *e.g.*, medical and military. Electromagnetic interference shielding is generally obtained by enclosing the device in a package made of electrically-conductive magnetic materials, which act as a barrier towards the EM radiation. [4-5] Usually, metallic materials *e.g.* magnesium-, [6] silver-, [7] and copper-based electronic packaging [8] are used to protect sensitive electronic components by reflecting most of the incident radiation. Alternatively, conductive polymer composites (CPCs) are attracting attention due to their mechanical flexibility, lightweight, corrosion resistance, and easy-processing at low cost. [5] Noteworthy, the EMI shielding effectiveness (EMI SE) of CPCs is mainly due to the EM wave absorption mechanism, which converts the incident electromagnetic energy into heat. [9] The energy conversion depends on the electrical conductivity and the electrical and magnetic polarization losses of the material. [9] The increase in temperature, due to energy conversion, might affect the functionality of the electronic components, reducing their lifetime. [9] In this case, CPCs obtained by using electrically and thermally conductive fillers, *i.e.* carbon nanotubes (CNTs), [10-14] graphene [15-19] or reduced graphene oxide (RGO), [20] MXene [21-26] represent valuable solutions to introduce a shielding layer able to protect the device and dissipate the heat simultaneously. In this context, CNT-Fe$_3$O$_4$@Ag/epoxy nanocomposites were designed, exhibiting EMI SE and thermal conductivity equal to 35 dB and 0.46 W m$^{-1}$ K$^{-1}$, respectively. [27] Graphene-based materials have also been used, *i.e.*, Graphene/RGO foam/epoxy, for the realization of EMI shielding nanocomposites, obtaining values for EMI SE up to 51 dB with thermal conductivity of 1.56 W m$^{-1}$ K$^{-1}$. [28]

Under specific conditions, *i.e.,* significant overheat of electronic devices, it may be useful to use EMI shielding materials exhibiting both in-plane electrical conductivity and through-plane thermal conductivity. In this case, such an anisotropic material will stop the incident EM radiation and will provide the necessary thermal protection of electronic devices through heat dissipation. In this context, anisotropic CPCs with high through-plane thermal conductivity, high in-plane electrical conductivity and low through-plane electrical conductivity, are suitable to meet the key specific requirements for the realisation of smart electronic packaging. Alongside with carbonaceous conductive fillers (*i.e.* CNTs and graphene), the hexagonal boron nitride (*h*BN) filler, being an electrical insulating and thermally conductive material, has recently attracted great attention as a suitable candidate to construct anisotropic materials through the layer-by-layer (LBL) approach. [29-31] In this regard, LBL (32 layers) polyethene/*h*BN/CNTs/graphite composites displayed a thermal conductivity of 1.45 W m$^{-1}$ K$^{-1}$. [32] Similarly, a silicon rubber/graphene/*h*BN composites (16 layers) exhibited a thermal conductivity of 1.25 W m$^{-1}$ K$^{-1}$, the electrical resistivity of 10$^{12}$ Ω cm in through-plane direction, and reached an EMI SE value of 40.67 dB with a filler content of ∼17.8 vol%. [33] It is worth mentioning that the composites with good electrical conductivity along the in-plane direction and high thermal conductivity in the through-plane direction can be used as electrically driven heaters with potential use as defoggers or defroster devices. [34] In these devices, laminated polymer/conductive filler/*h*BN composites can be heated applying a bias voltage, removing ice and protecting the device at low temperatures operations. Simultaneously, the heat produced by the device is dissipated. [35] However, the production of composites with LBL structure requires a large amount of fillers to achieve the required combination of thermal conductivity, Joule effect and EMI shielding properties. [36, 37] Therefore, the designing and preparation of composites with a more efficient LBL segregated filler network may represent a viable approach for reducing the filler content and improving both the thermal and electrical conductivity of layered structure polymer composites. [36-39]

[a] School of Materials Science and Engineering, Liaocheng University, Liaocheng 252000, China
[b] Graphene Labs, Istituto Italiano di Tecnologia, via Morego 30, 16163 Genoa, Italy
[c] Dipartimento di Chimica e Chimica Industriale, Università degli Studi di Genova, via Dodecaneso 31, 16146 Genoa, Italy
[d] Institute of Polymers, Composites and Biomaterials, National Research Council, P.le Fermi, 1-80055 Portici, Naples, Italy. E-mail: marino.lavorgna@cnr.it
[e] State Key Laboratory of Polymer Materials Engineering, Polymer Research Institute, Sichuan University, Chengdu 610065, China. E-mail: xiahs@scu.edu.cn
[f] Guangxi Key Laboratory of Calcium Carbonate Resources Comprehensive Utilization, Hezhou University, Hezhou, China
[g] BeDimensional S.p.a., Via Albisola 121, Genova 16163, Italy. E-mail: a.delrio@bedimensional.it

In previous reports, the segregated carbonaceous filler network was constructed through a rubber-based latex mixing approach. [4, 20, 36, 39] In details, the latex particles force the filler into the excluded-volume in between the particles forming a continuous network, thereby reducing the volume of filler necessary to achieve the electrical percolation (less than 0.62 vol% as compared to 4.62 vol% when the filler is randomly dispersed). [39, 40] Besides the high-electrical and -thermal conductivity at the lowest concentration needed to reach the percolation threshold, the additional advantages of this method are the ease of mixing and its eco-sustainability due to the use of water as the solvent. Composites with filler segregated network, such as $Fe_3O_4$@RGO/rubber, [20] CNT/rubber, [41] $Fe_3O_4$@graphene/poly(methyl methacrylate), [42] polystyrene/BN, [43] ultra-high-molecular-weight-polyethylene/BN, [44] polypropylene/graphite, [45] were all prepared by filler self-assembling and latex method for EMI shielding and thermally conductive applications.

Natural rubber latex obtained from the *Hevea brasiliensis* tree is an environmentally friendly and low-cost material, which can be easily processed with well-established technologies.[46] These advantages, combined with its excellent mechanical properties, make the natural rubber an ideal candidate to realize composites useful for many applications, *e.g.*, tires, seals or shock absorptions. In this work, natural rubber-based composites exhibiting a layered structure have been prepared by combining the mixing and the vacuum-assisted filtration methods. Specifically, rubber-based layered composites with CNTs and *h*BN (RCB), obtained by alternating layers made of natural rubber (NR)/CNTs and NR/*h*BN exhibit high electrical conductivity (1 S cm$^{-1}$) and high electrical insulation in the in-plane direction and thermal conductivity in the through-plane direction (0.25 W m$^{-1}$K$^{-1}$). When EM waves enter into the composites from NR/*h*BN layer, the EMI SE of RCB composite with layers containing *h*BN at 16 phr (∼13.8 wt%) and CNTs at 8 phr (∼7.4 wt%), with a total thickness of 1.4 mm, is 31.38±0.2 dB. At the same time, the heat energy converted from EMI energy can be dissipated into the environment because of its high through-plane thermal conductivity (0.25 W m$^{-1}$ K$^{-1}$). This rubber-based anisotropic material, characterized by layered segregated filler morphology, represents an important step toward real applications of anisotropic materials, exhibiting a combination of high EMI shielding, thermal conductivity and heat-dissipation capability.

## Experimental

**Material**

Pre-vulcanized natural rubber latex (HMR 10, solid content: 60.5wt%) was supplied by Synthomer, UK. Hexagonal boron nitride, *h*BN (CAS: 10043-11-5) was obtained from Alfa Aesar. Carbon nanotube (NC 7000, diameter: 10 nm, length: 1.5 μm, density: 1.75 g cm$^{-3}$) were purchased from Nanocyl S.A., Belgium. Graphite, Mesh 100, was purchased by Sigma Aldrich. Cetyltrimethylammonium bromide (CTAB), as a surfactant, was obtained from Sigma Chemicals Company. All agents are used without further purification.

**Exfoliation of Boron Nitride**

A mixture of the bulk layered boron nitride (100 g) and the solvent (10 L of NMP, Sigma Aldrich) is prepared and exfoliated using a Wet Jet Mill (JN100, Joko, Japan). The Wet Jet Mill apparatus consists in a hydraulic mechanism and a piston, which supplies a pressure (250 MPa) pushing the boron nitride and NMP mixture into the processor , in which jet streams, high shear rates and cavitation are generated. [44,45] The mixture of bulk boron nitride and NMP is placed in a container and processed using a 0.10 mm nozzle diameter. The sample is totally processed, setting the piston passes at 1000. The process is repeated 10 times to guarantee the processing of the whole sample. The processed sample is then collected in a second container. The flake lateral sizes and thickness are reported in Ref 47 indicating a lateral size of 360 nm and thickness mode of 2.4 nm. The full characterisation of the $h$BN flakes is reported in the supplementary **Figure S1a** to **c**.

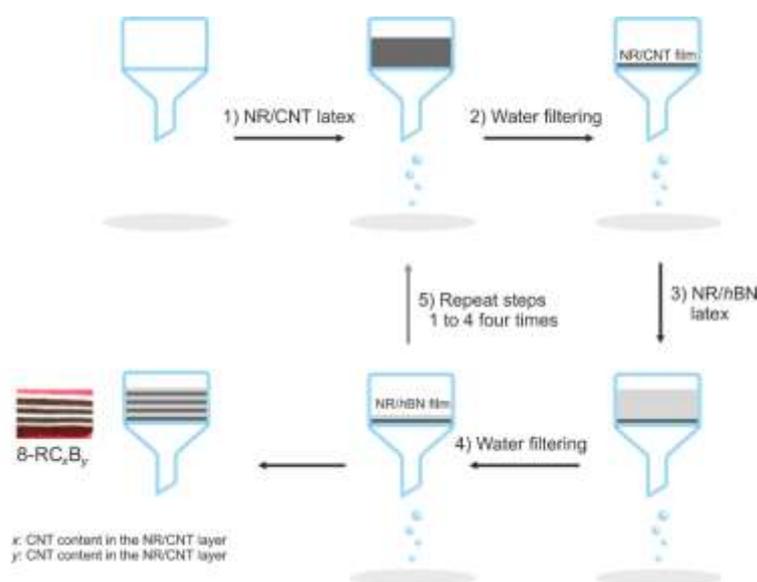

**Figure 1**. Schematic illustration of the preparation process of RCB composites with a layered structure. The picture shows the 8-RC$_8$B$_8$ composite. The same process was adopted to produce some layered structures by using graphene as alternative to carbon nanotubes.

**Exfoliation of Graphite**

A mixture of the Graphite (100 g) and the solvent (10 L of NMP) is prepared and exfoliated using a Wet Jet Mill (JN100, Joko, Japan). [47-49] The mixture of Graphite and NMP is placed in a container and processed using consecutively nozzles with diameters of 0.3, 0.15 and 0.10 mm. The process is repeated 2 times using the 0.10 mm nozzle. The processed sample is then collected in a second container. The flake lateral sizes and thickness are reported in Ref 47 indicating a lateral size of 460 nm and thickness mode of 1.6 nm. The full characterisation of the few-layers graphene flakes is reported in the supplementary **Figure S1d** to **f**.

**Preparation of the composites with a layered-structure**

Measured amounts of CNTs and CTAB with a CNTs/CTAB weight ratio of 1:1 were dispersed in water by using an ultrasound probe (UP200S Hielscher sonic bath, the temperature was controlled with an ice-water mixture in a beaker, cycle: 0.5, amplitude: 80%) for 30 min. Pre-vulcanized natural rubber latex was added into the CNTs dispersion and further sonicated for 30 min, realizing NR/CNTs latex. NR/$h$BN latex was obtained through the same procedure. The experimental formulation of latex composites is shown in **Table S1**.

NR/CNTs latex (6 ml) was placed in a Buchner funnel and filtrated by vacuum-assisted filtration to deposit the first composite layer on the paper filter. The CNTs, initially assembled onto the surface of rubber latex particles, gives rise to an LBL morphology in the samples, **Figure 1**. Subsequently, NR/$h$BN latex (6 ml) was poured on the first layer made by NR/CNTs (black NR/CNTs layer) and water was removed by filtration, forming the second layer (white NR/$h$BN layer). These steps were repeated until the desired number of layers was achieved. Finally, the films were dried at 60 °C for 6 h and then heated at 150 °C for 30 min in the vacuum oven to vulcanize the natural rubber. The obtained films were designated as $z$-RC$_x$B$_y$, where $z$ represents the total number of layers, $x$ and $y$ represent the CNTs and $h$BN content in the layers, respectively. For a comparative purpose, NR/graphene/$h$BN composites were prepared by using the same process and replacing CNTs with graphene and $h$BN filler.

**Characterization**

Transmission Electron Microscopy (TEM) images of RCB composites were carried out by using an FEI Tecnai G$^2$ F20 S-TWIN transmission electron microscope, operating at an accelerating voltage of 200 kV. The samples were sliced at room temperature (70-80 nm thickness) using a cryo-microtomed Leica EM UC6 equipment. The cut films were collected and supported on copper grids for observation.

Scanning electron microscopy (SEM) was conducted using a Zeiss Ultra 55 apparatus (Jena, Germany) with a detector operating at an accelerating voltage of 10 kV. Specimens were sputter-coated with 10 nm gold before measurement using an EMS550X sputter coating system. In order to evaluate the effect of bending cycles at the interface between two layers, samples are bend at 90º for 1000 cycles and subsequently SEM analysis were performed. Moreover, SEM investigations of tensile fracture surfaces of the samples were also executed. The chemical composition of the films was analysed by X-ray energy dispersion spectroscopy (EDS, Oxford Instruments, coupled to the SEM microscopy), by applying 3 keV acceleration voltage under 8.5 mm working distance.

The thermal conductivity ($\kappa$, W m$^{-1}$ K$^{-1}$) of RCB composites was calculated from the thermal diffusivity ($\alpha$, mm$^2$ s$^{-1}$), specific heat capacity ($C_P$, J g$^{-1}$ °C$^{-1}$), and the density ($\rho$, g cm$^{-3}$), using the relationship $\kappa = \alpha \times C_p \times \rho$.[32]

The thermal diffusivity of RCB discs (diameter of 12.7 mm) was measured at 25 °C using a laser-flash apparatus (LFA 467, Netzsch Instruments, Inc.).

TA Discovery Differential Scanning Calorimetry, DSC instrument was used to estimate the $Cp$ of RCBs with different filler content. All samples were heated from 0 to 100 ºC, then cooled down to 0 °C and finally re-heated up to 100 °C at the rate of 10 °C min$^{-1}$. The DSC curve of the second heating ramp was used in order to obtain the $Cp$ value at 25 °C. Density was calculated as the ratio of mass to volume of the sample. The mass was measured by an analytical balance with a precision of 0.001 g, whereas the sample volume was determined by the geometrical shape measuring the dimensions by using a calliper with an accuracy of 0.01 mm.

The electrical conductivity of all composites was measured by a two-point measurement (Keithley 2400 picoammeter). Rectangular samples (10×5×1.4 mm$^3$) were cut and coated with a silver epoxy paste (Shenzhen Sinwe New Material Co. Ltd) to assure good electrical contact between picoammeter electrodes and the sample surface.

Thermogravimetric analysis (TGA) was carried out using a TGA Q500 (TA Instrument) in an air atmosphere to characterize the thermal stability of samples. The samples were heated from 30 ºC to 700 ºC at a heating rate of 10 ºC min$^{-1}$.

Fourier-Transform infrared spectroscopy (FTIR) spectra were recorded at room temperature by using a FT-IR spectrometer (model Frontier Dual Ranger, PerkinElmer, USA) in attenuated total reflectance (ATR) mode from 650-4000 cm$^{-1}$. ATR spectra were recorded at 4 cm$^{-1}$ resolution, and the reported results are the average of 32 scans. All the spectra were background-corrected.

To evaluate the electric heating behaviour, the silver-coated samples (10×5×1.4 mm$^3$) were connected by copper wires to the DC power supply. An IR camera (FLIR A35) was used to record the temperature of samples surface at the different input voltages. The DC power supply was interrupted when the surface temperature reached 170 °C.

The EMI SE of the samples (thickness of 1.4 mm) was evaluated by using an Agilent N5247A vector network analyser in a transmission-reflection mode. The scattering parameters ($S_{11}$ and $S_{21}$) in the frequency range between 8.2 and 12.4 GHz (X-Band) were recorded. From the $S_{11}$ and $S_{21}$ scattering parameters, the power coefficients of reflectivity ($R$), transmissivity ($T$), and absorptivity ($A$) is obtained using the following equations:[50]

$$R = |S_{11}|^2 \qquad (1)$$

$$T = |S_{21}|^2 \qquad (2)$$

$$A = 1 - R - T \qquad (3)$$

Therefore, the effective absorbance ($A_{eff}$) can be described as follows:[20]

$$A_{eff} = A/(A+T) \qquad (4)$$

The total EMI SE ($SE_T$), defined as the logarithmic ratio of incoming ($P_{in}$) to outgoing power ($P_{out}$) of electromagnetic radiation was also calculated, according to equation (5): [51]

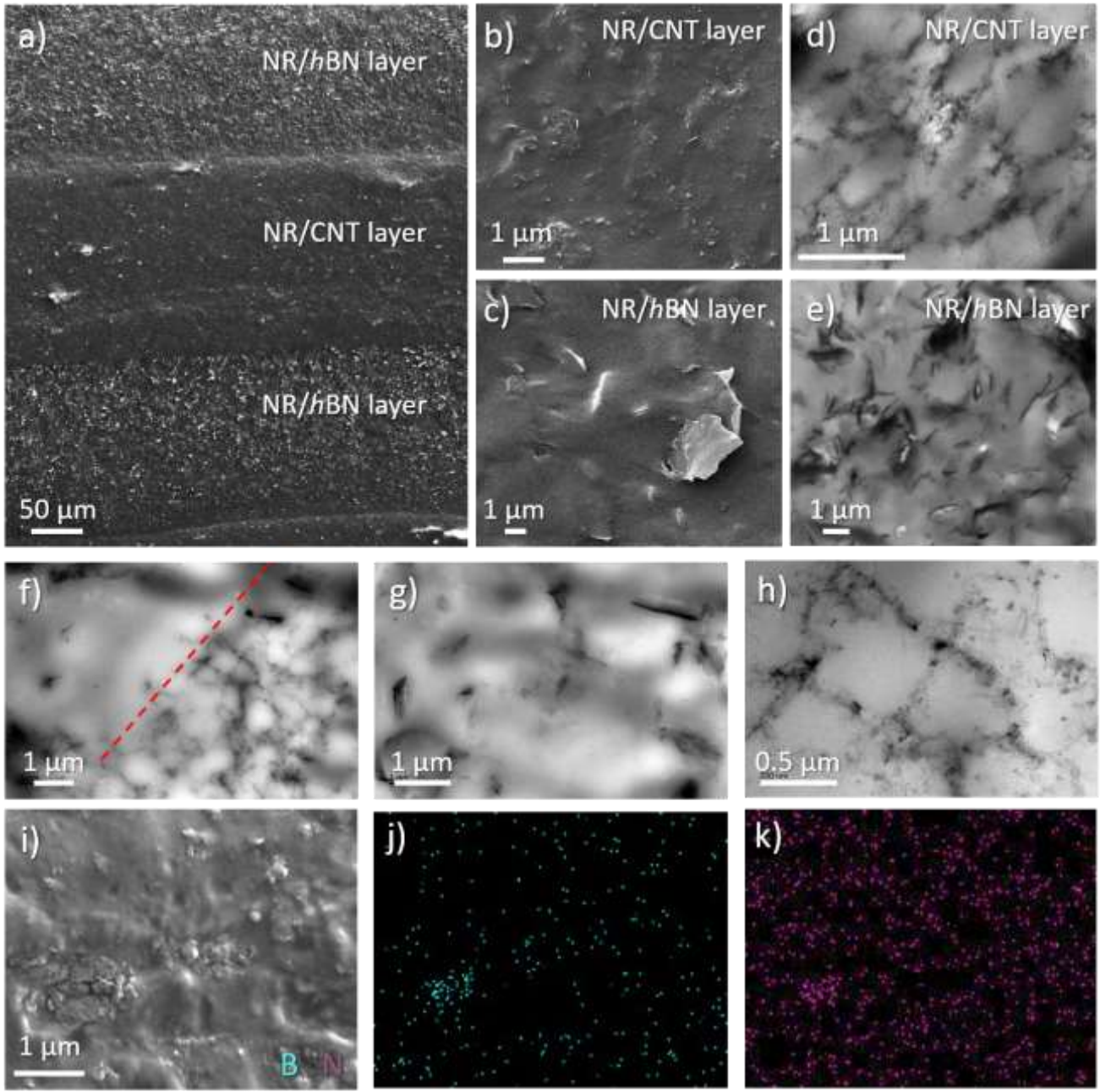

**Figure 2**. SEM images of section surface of 8-RC$_8$B$_{12}$ composite (a), NR/CNTs layer (b) and NR/$h$BN layer (c). TEM images of NR/CNTs layer (d) and NR/$h$BN layer (e). A TEM image of the interface zone between two joined layers (f), NR/$h$BN layer (h) and NR/CNTs layer (j) for 8-RC$_8$B$_8$ composites. SEM image (i), B mapping (j) and N mapping (k) of 8-RC$_8$B$_8$ composites.

$$SE_T = -10 lg\left(\frac{P_{out}}{P_{in}}\right) = SE_R + SE_A + SE_M \qquad (5)$$

in which $SE_A$, $SE_R$, and $SE_M$ are the absorption, reflection, and multiple reflections shielding, respectively. The $SE_R$ and the sum $SE_A$ and $SE_M$ can be obtained from equations (6) and (7): [50]

$$SE_R = -10 lg(1-R) \quad (6)$$

$$SE_A + SE_M = -10 lg(\frac{T}{1-R}) \quad (7)$$

in which $T$ and $R$ are the transmissivity and reflectivity, respectively, as defined by equations (1) and (4).

## Results and discussion

### Segregated layered morphology and structure of RCB composites

The layered structure of 8-RC$_8$B$_8$ sample is evident in the picture in **Figure 1**, in which the black layers correspond to the NR/CNT composite and the white ones to the NR/$h$BN layers. Each layer displays a uniform thickness (about 190 µm ± 19 µm), as shown in **Figure S2a**, which is attributed to the reliability and robustness of the vacuum-assisted filtration process. The NR/CNT layer is well joined to the NR/$h$BN layer, being a result of a wet-wet deposition step, and a sharp interface without cracks between the two layers is observed (**Figure S2b**). A merged interface between adjacent layers is observed for the sample 8-RC$_8$B$_{12}$ all over its longitudinal section. Furthermore, the same merged interfaces can be observed for the 8-RC$_2$B$_8$ sample submitted to bending at 90° for 1000 cycles (**Figure S2c**) and after tensile deformation until break (**Figure S2d**). These results prove that the interface between adjacent layers is sufficiently strong to resist to stress and preserve the layered structure from delamination. The boundary between layers is also observable by SEM and TEM, shown, in **Figure 2a** and **Figure 2f** respectively. For clarity reasons, the red dotted line is used as eye-guide (**Figure 2f**). The presence of such interfaces suggests that the NR latex containing the filler does not penetrate the compact layer, already deposited during the filtration process. The segregated morphology of $h$BN nanoparticles and CNTs in the composite layer of 8-RC$_8$B$_{12}$ sample is shown in **Figure 2b** to **e**. The filler morphology of composite 8-RC$_8$B$_8$ is shown in **Figure 2f-h**. Compared to 8-RC$_8$B$_8$ sample, the 8-RC$_8$B$_{12}$ sample exhibits a better defined segregated $h$BN morphology due to the presence of a larger amount of $h$BN filler. The segregated networks demonstrate that the fillers remain encapsulated in the rubber particles during the latex mixing stage, forming a three-dimensional network during filtration when the filler is forced between the latex particles. To highlight the structure of the segregated $h$BN network, SEM and Energy-dispersive X-ray spectroscopy (EDS, including the contribution of B and N elements in the NR/$h$BN layer of 8-RC$_8$B$_8$ composite) are shown in **Figure 2i-k** and **Figure S3**. It is noted that B and N element dots, which identify the $h$BN platelets.

### Through-plane thermal conductivity of RCB composites

The thermal conductivity in the through-plane direction of the RCB composites increases with the filler content, as illustrated in **Figure 3a**. The value for the 8-RC$_8$B$_{12}$ composite thermal conductivity is 0.25 W m$^{-1}$ K$^{-1}$, which is about 50% higher than that of pristine NR (0.17 W m$^{-1}$ K$^{-1}$). The higher thermal conductivity can be ascribed to both the high thermal conductivity of CNTs (3000 W m$^{-1}$ K$^{-1}$) [52] and $h$BN (600 W m$^{-1}$ K$^{-1}$) [53] nanoparticles (which form the segregated network between rubber particles and provide effective heat conductive paths [37, 54]), and to the healed interface between the adjacent layers (which make contact between adjacent layers and provide an effective heat transport medium in the through-plane direction).

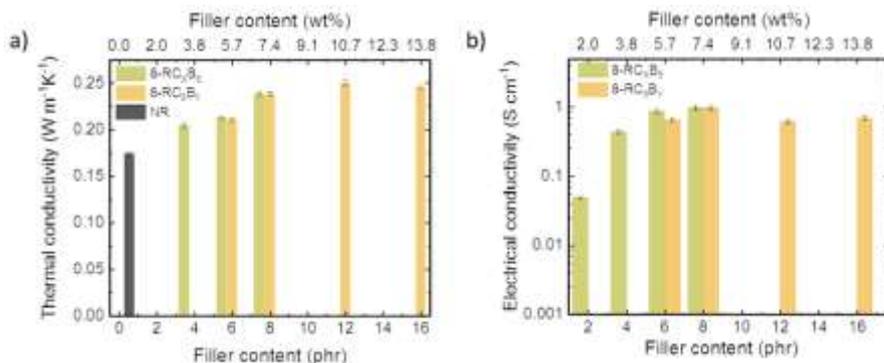

**Figure 3**. a) Thermal conductivity of RCB composites in the through-plane direction. b) The electrical conductivity of 8-RC$_x$B$_y$ composites.

It is worth noting that the total or even partial substitution of CNTs with graphene brings a 100% increment the thermal conductivity compared to the pristine NR sample, the results are summarised in **Table S2** for samples 8-R(GE$_4$C$_4$)B$_8$ and 8-RGE$_8$B$_8$, (exhibiting a thermal conductivity of 0.35 W m$^{-1}$ K$^{-1}$, which is ascribed to the presence of graphene flakes (thermal conductivity of 5000 W m$^{-1}$ K$^{-1}$).[55] **Figure S3** shows the variation of the heat capacity, $C_p$ for the RCB composites at 25 °C with the increase of the filler content. In particular, the $C_p$ decreases with the addition of CNTs or $h$BN filler due to the intrinsically lower $C_p$ of CNTs (0.65 J g$^{-1}$ K$^{-1}$)[56] and $h$BNs (0.78 J g$^{-1}$ K$^{-1}$)[57] compared to the pristine NR (2.00 J g$^{-1}$ K$^{-1}$). Taking into account that the $C_p$ decreases with filler content and that the density of the composite does not change significantly, it is possible to confirm that the thermal diffusivity, $\alpha$, is the key parameter to control the thermal conductivity of LBL composites. By increasing the filler content, the thermal diffusivity increases as a consequence of the thermal phonon exchange due to the densification of the three-dimensional filler network and increase of overlapping between particles, as shown in the section related to the morphology.[58]

**The in-plane electrical conductivity of RCB composites**

The electrical conductivity of the RCB composites affects greatly their electro-heating behaviour and EMI shielding performance. By cutting-off the conductive path with the insulated NR/$h$BN layer, the RCB composites become electrical insulators in the through-plane direction (**Figure S5a**), while RCB composites are conductive in the in-plane direction (**Figure S5b**). The in-plane electrical conductivity data of 8-RC$_x$B$_y$ composites are displayed in **Figure 3b**, showing that the electrical conductivity of 8-RC$_x$B$_8$ composites increases with increasing the CNTs content, which is ascribed to the formation of a more effective conductive network.[39] In particular, the electrical conductivity of 8-RC$_8$B$_8$ is 0.98 S cm$^{-1}$, which is similar with the values exhibited by some of our previous NR/CNT composites (CNT content: 8 phr, 0.30 S cm$^{-1}$)[5] and by NR/CNT composites (CNTs content: 10 wt%, 1.31 S cm$^{-1}$) published by Jia et al.[41]. The results suggest that the effective CNT segregated network is preserved during the filtration process, as shown in **Figure 2**. From the data reported in **Table S2**, it is possible to see that the electrical conductivity of 8-RC$_8$B$_8$ (0.98 S cm$^{-1}$) is higher than the ones of the samples 8-R(graphene$_4$C$_4$)B$_8$ (0.234 S cm$^{-1}$) and 8-R(graphene$_8$)B$_8$ (3.98×10$^{-5}$ S cm$^{-1}$). The satisfying electrical conductivity of the RCB composites is also demonstrated by the gleaming of a Light Emitting Diode (LED) bulb connected to the battery through an electrical circuit that includes the 8-RC$_8$B$_{12}$ composite (**Figure S5c**). Furthermore, it is noted that the electrical conductivity of RCB composites decreases with increasing the $h$BN content from 8 phr to 16 phr. This behaviour can be ascribed to the effect of accumulation of CTAB in the layers with CNTs, which realizes during the filtration. Likely, the surfactant, coming from the layers with $h$BN, deposits onto the surface of CNTs leading to an increase of the distance between them and reducing, consequently, the effect of electron tunnelling transport.[55] This hypothesis is supported by FTIR results for NR/CNT layers and NR/$h$BN layers which show an accumulation of CTAB in the CNTs layers of composite samples with higher contents of $h$BN (**Figure S6**).

**EMI shielding property of RCB composites**

The EMI SE behaviour of materials depends on the reflection and absorption of the incident EM waves at the interfaces between both air/material, between different materials and within the material.[20] In particular, the reflection contribution $SE_R$ is related to the impedance mismatch between the air and the material,[20] in which the absorption, $SE_A$ depends on the conduction, polarization and magnetic losses of the adsorbing materials.[33] The case of multiple reflections, $SE_M$ is often neglected for isotropic materials,[20] *i.e.* when the thickness of the sample is much bigger than the depth skin (which represents the penetration distance of radiation in the materials at which the intensity transmitted is reduced to 1/e of its initial value). In our cases, multiple reflections (*i.e.*, air/material, material/air and material based on CNT/materials based on $h$BN) cannot be neglected since each layer is about 190 µm ± 19 µm (as shown by SEM images, **Figure S2a**).

Since the impedance mismatch between the air and the NR/CNT layer is different from that between the air and the NR/$h$BN layer, the EMI shielding property of RCB composites (1.4 mm in thickness) were measured with respect to the incident directions of EM wave, *i.e.,* a) EM waves entering into the sample through the NR/CNTs layer and b) through the NR/$h$BN layer, as shown in **Figure 4**. In particular, the EMI SE were determined in the range of 8.2-12.4 GHz, which represents the working frequencies band for transmission radar communication and for most electronic devices, *i.e.* wireless devices such as cell phone, tablets, televisions, computers, supercomputers.[1] The data show that the EMI SE of all samples increases with the frequency, regardless of the direction of the incident waves, as shown in **Figure 4a** and **b**. Moreover, although the density of RCB composites does not change with the filler content (**Figure S7**) their EMI SE does. The increase of EMI SE is attributed to the increment of free-charge carriers and electric dipoles, together with the enhanced conductive network. [42] In particular, the EMI SE of 8-RC$_4$B$_8$ is 22.68±0.2 dB at 10.3 GHz, when the EM waves penetrate into the samples from the NR/CNT layer. This value satisfies the commercial EMI shielding requirement (20 dB) for shielding materials. [20] Moreover, by increasing the $h$BN content, the 8-RC$_8$B$_y$ composites show a further increment of EMI SE values, *i.e.,* the EMI SE values for 8-RC$_8$B$_6$ and 8-RC$_8$B$_{16}$ are 26.29±0.2 and 29.93±0.2 dB, respectively, at 10.3 GHz. The increment of EMI SE is ascribed to a denser and more interconnected $h$BN network, allowing greater availability of pathways for the movement of charges from the CNT layer to the $h$BN layer through a hopping mechanism. [60] However, it is also possible that an increase in the $h$BN content brings a significant impedance mismatch between the NR/$h$BN and the NR/CNT layers, thereby enhancing the internal reflections of EM. It is worth mentioning that the RCB composites exhibit higher EMI SE values when the EM waves enter into the samples from NR/$h$BN layer (**Figure 4b**). In fact, the EMI SE values of 8-RC$_8$B$_8$ and 8-RC$_8$B$_{16}$ are 30.49±0.2 and 32.52±0.2 dB, respectively, at 10.3 GHz. To understand the mechanisms involved in the electromagnetic shielding effect of RCB composites, the different contributions $SE_A+SE_M$, $SE_R$ and $SE_T$ for 8-RC$_x$B$_8$ and 8-RC$_8$B$_y$ composites, have been included in **Figure 4c** and **Figure 4d**. The values of $SE_A+SE_M$ and $SE_T$ for 8-RC$_8$B$_y$ and 8-RC$_x$B$_8$ composites increase with the filler content, while $SE_R$ does not change significantly. In particular, the $SE_R$ value (~1 dB) when the waves traverse the sample from the NR/$h$BN layer is lower than that (~5 dB) when the waves traverse the sample from the NR/CNTs layer. The phenomenon is attributed to the impedance mismatch between air and NR/$h$BN layer which is lower compared to the mismatch between air and NR/CNT layer. Furthermore, for all 8-RC$_x$B$_y$ composites, the contribution of $SE_A+SE_M$ to the total EMI SE is more than 98 %, as shown in **Figure S8**. These results confirm that the EM wave absorption and multiple reflections are the key mechanisms mainly contributing to the EMI SE of 8-RC$_x$B$_y$ composites, which implies that the layered structure and the filler segregated morphology improve the adsorption of EM rather than reflect it into the surroundings. However, the RCB composites exhibit the highest $SE_A+SE_M$, and lowest $SE_R$ contributions when the EM waves enter from NR/$h$BN layer and leave from NR/CNT layer. In particular, the data in **Figure 4c** and **4d** show that the $SE_A+SE_M$ of RCB composites measured when EM waves are incident from NR/$h$BN, is always higher than $SE_T$ obtained when EM waves enter from NR/CNT layer at 10.3 GHz. Thus, the results confirm that the best configuration to increase the shielding effectiveness is achieved when the EM waves enter the sample from NR/$h$BN layer and leave from NR/CNT layers. To further evaluate the EMI shielding performance it is useful to use the normalized EMI SE values (*i.e.* EMI SE divided by sample thickness). The specific EMI SE value of the 8-RC$_8$B$_{12}$ composite (1.4 mm thickness) is equal to 20.62±0.14 dB mm$^{-1}$ when the EM wave enters from NR/CNT layer, while it is equal to 22.41±0.14 dB mm$^{-1}$ when the EM wave enters from NR/$h$BN layer at 10.3 GHz. Both these EMI values are higher than some of the earlier published values for rubber-based EMI materials (**Table S3**).

The skin depth parameter ($\delta$), is a quantitative measure of the materials' shielding ability. [20] Moreover, the skin depth is inversely proportional to the $SE_A$ value of materials at a fixed thickness, $d$. [61] The relationship between $SE_A$ and skin depth is given by equation (8): [42]

$$SE_A = 20\log(e^{d/\delta}) = 8.686(d/\delta) \qquad (8)$$

The skin depth depends on the frequency according to equation (9): [42]

$$\delta = 1/(\pi f \mu \sigma)^{1/2} \qquad (9)$$

where $f$ is the frequency, $\sigma$ is the electrical conductivity of the overall anisotropic composite, and $\mu$ is the magnetic permeability of material ($\mu=\mu_o\mu_r$ where $\mu_o=4\pi \times 10^{-7}$ H m$^{-1}$ and $\mu_r$ is the material's relative magnetic permeability, and $\mu_r=1$ for the nonmagnetic composites). Neglecting the multiple reflections from each layer, the skin depth of the samples with more than 4 phr (~3.8 wt%) CNT is estimated to be less than 0.7 mm, see **Figure 4e**. Since this skin depth value (around 0.7 mm) corresponds to the total thickness of the four NR/CNT layers, it can be deduced that the RCB composites display excellent EMI shielding properties when the CNT concentration of the NR/CNT layers is greater than 4 phr.

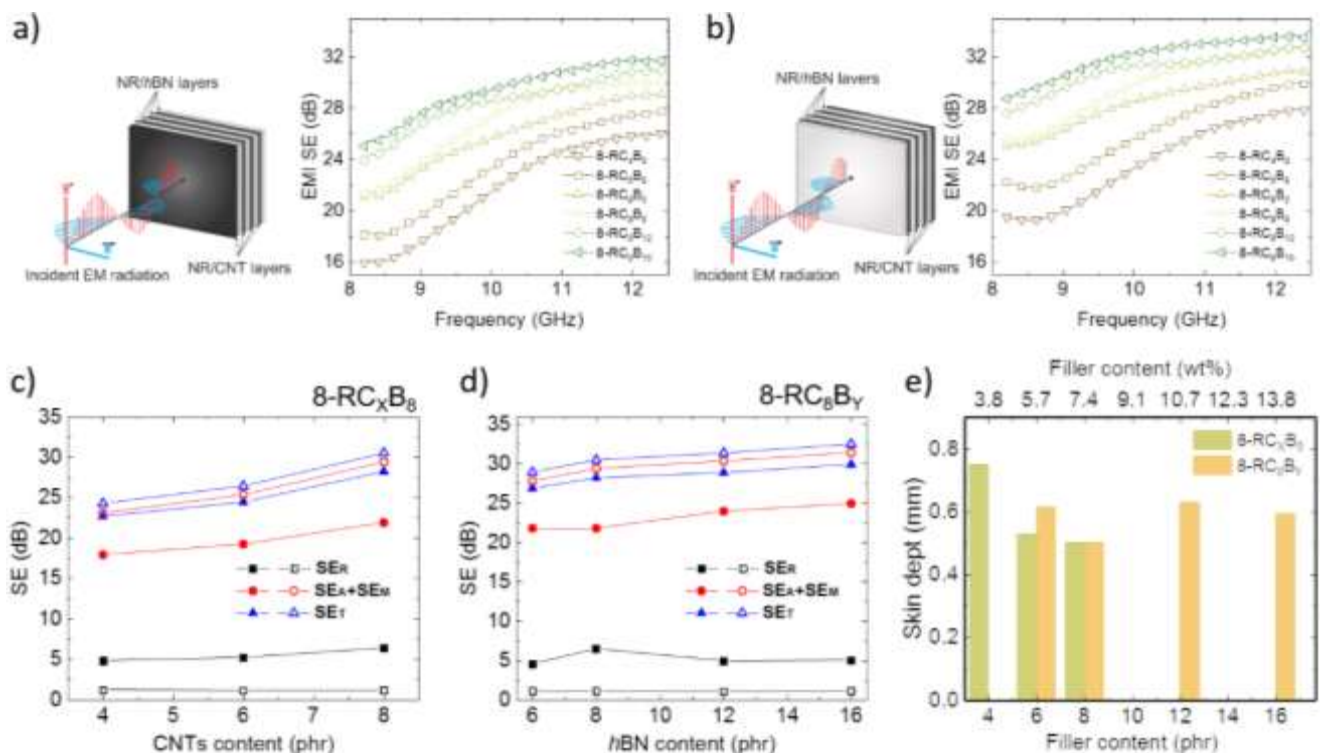

**Figure 4**. a) Electromagnetic waves are incident from the NR/CNT layer; EMI SE of RCB composites with different filler content as a function of frequency when EM waves penetrating the samples from NR/CNT layer; b) EM waves enter from NR/$h$BN layer. and EMI SE of RCB composites with different filler content as a function of frequency when EM waves penetrate the samples from NR/$h$BN layer; c) Shielding by reflection, absorption and multiple reflections, and total shielding of 8-RC$_x$B$_8$ composites at 10.3 GHz (solid symbol: EM waves enter from NR/CNT layer; hollow symbol: EM waves enter from NR/$h$BN layer). d) Shielding by reflection, absorption and multiple reflections, and total shielding of 8-RC$_8$B$_y$ composites at 10.3 GHz (solid symbol: EM waves enter from NR/CNT layer; hollow symbol: EM waves enter from NR/$h$BN layer). e) The skin depth of the 8-RC$_x$B$_8$ and 8-RC$_8$B$_y$ composites, in the hypothesis the multiple reflections are neglected. The sample thickness is 1.4 mm. Error bar of Agilent N5247A vector network analyser was 0.2 dB.

**Electric heating behaviour of RCB composites**

Materials exhibiting significant electrical conductivity are often used as electric heaters. Electrical energy dissipated by the Joule effect was monitored by infrared imaging of the NR/$h$BN surface (**Figure S9**). During the application of the voltage (5 V), the samples exhibited a uniform temperature distribution (inset imagine of **Figure 5a**). The time-dependent temperature profiles of 8-RC$_4$B$_8$ composites at the different voltages (from 2.5 V to 7.5 V) are shown in **Figure 5a**. The surface temperature of 8-RC$_4$B$_8$ increases to 170 °C in 30 s applying an electric potential of 7.5 V. When the voltage is reduced to 5 V and 2.5 V the equilibrium surface temperature of the 8-RC$_4$B$_8$ composites drastically decreased to 165.4 °C and 47.8 °C, respectively, suggesting that there is sharp non-linearity (switching effect) in the heat-voltage. [61] These results confirm that the sample 8-RC$_4$B$_8$ can be used as heat dissipater by applying lower heat-voltage as compared to the ones previously reported for graphene/rubber, [34] *i.e.* 10 V and for poly(3,4-ethylene dioxythiophene) textiles, [62] *i.e.* 6 V. The heat-voltage is similar to the value found for both MXene/polyester composite, [63] *i.e.*, 2 V and graphene reports, [64] *i.e.*, 3.2 V. These low-trigger voltage values not only ensure human body safety when the devices are used as textile heaters but make it also possible to be powered by portable batteries. [63] The input voltage of 2.5 V was applied to all 8-RC$_x$B$_y$ composites for investigating the effect of filler contents on their electric heating behaviour. The results are shown in **Figure 5b**.

It is worth noting that the stationary temperature of all samples is reached faster with the increase of the CNT content. The steady-state temperature changes from 32.1 °C to 96.8 °C when CNT content increases from 2 phr (~2 wt%) to 8 phr, which is attributed to the increase in the electrical conductivity of 8-RC$_x$B$_8$. It is also significant to note that the surface temperature of 8-RC$_8$B$_y$ composites increases from 74.6 °C to 103.3 °C when $h$BN content increases from 6 phr (5.66 wt%) to 12 phr. That is ascribed to the fact that the thermal conductivity increases with $h$BN content whereas the $Cp$ decreases, suggesting a more significant Joule heating effect from inner layer to surface layer. Based on data reported in **Table S4**, the RCB composites display better electric heating behaviour than epoxy/graphene films,[65,66] epoxy/graphene/CNT films[66] and polydimethylsiloxane/CNT/graphene films [67] with the same sample size (length and wideness). The thermal stability of electric heating materials affects the temperature range wherein the samples can operate. Based on **Figure S10**, RCB composites can be safely used below 200 °C, which represents the temperature onset for thermal degradation of samples as determined by TGA in air atmosphere.

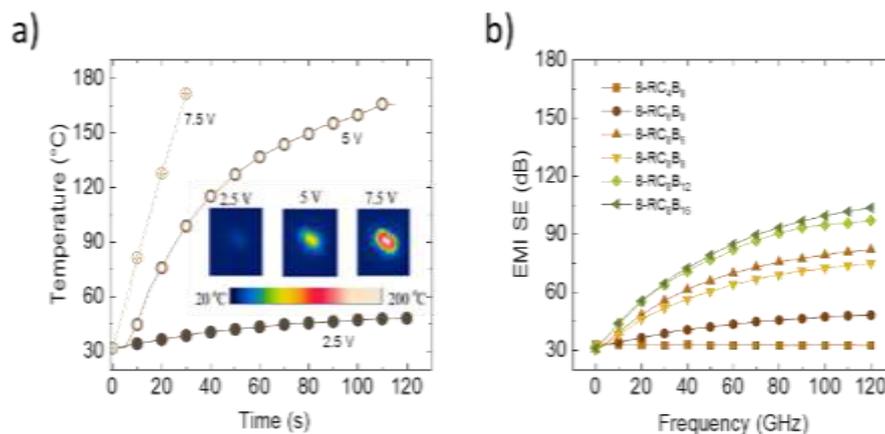

**Figure 5.** a) The time-dependent temperature profiles of 8-RC$_4$B$_8$ composites for the different input voltage. The inset picture is IR image of 8-RC$_4$B$_8$ composites surface after inputting voltage for 30 s. b) the effect of filler content on the electro heating behaviour with an input voltage of 2.5 V.

## Conclusions

Layered composites containing carbon nanotubes and boron nitride flakes fillers with a segregated morphology were produced by combining the latex mixing and vacuum-assisted filtration methods. These composites have been investigated as potential candidates for both electromagnetic interference (EMI) shielding application in the 8.4-12.4 GHz range and heater dissipaters. The results confirm, as compared to the state-of-the-art, that the segregated morphology of the fillers and the anisotropic structure built-up by alternating conductive and insulating layers. Enableing the realization of effective multifunctional EMI shielding materials, by exploiting ease and reliable methods. In the through-plane direction, the layered composites, *i.e.,* the 8 phr of carbon nanotubes and 12 phr of boron nitride flakes (8-RC$_8$B$_{12}$), display a thermal conductivity up to 0.25 W m$^{-1}$ K$^{-1}$, electrical insulation and an EMI shielding of 32.52±0.2 dB at 10.3 GHz. In the in-plane direction, the layered composites exhibit an electrical conductivity up to 0.98 S cm$^{-1}$, and electric heating behaviour. In particular, the thermal conductivity and specific EMI shielding efficiency of 8-RC$_8$B$_{12}$ composites (containing four layers with carbon nanotubes at 8 phr and four layers with boron nitride flakes at 12 phr) are respectively 0.25 W m$^{-1}$ K$^{-1}$ and 22.41±0.14 dB mm$^{-1}$ (EM waves enter from NR/ boron nitride layer). The surface of the same material reaches about 103°C with an input voltage of 2.5V. The present work opens the way to the design of next-generation multifunctional EMI shielding materials for possible use as "smart textiles" able to shield electronic device, capable of heating as well as dissipating heat emanating from external devices.

## Conflicts of interest

There are no conflicts to declare.

## Acknowledgements

This work was supported by the Joint Laboratory for Graphene-based Multifunctional Polymer Nanocomposites founded by CNR (Joint Labs Call 2015 - 2018), by the Bilateral project GINSENG between NSFC (China) and MAECI (Italy) (2018 - 2020), by Natural Science Foundation of Shandong Province (ZR2019QEM009), by opening project of Guangxi key laboratory of calcium carbonate resources comprehensive utilization (HZXYKFKT201803), by the special foundation for innovation-driven development of Hezhou (PT0710004) and the European Union's Horizon 2020 research and innovation program under grant agreement No. 785219-GrapheneCore2.

## Notes and references

1  P. M. Mariappan, D. R. Raghavan, S. H. E. Abdel Aleem and A. F. Zobaa, *J. Adv. Res.,* 2016, **7**, 727.
2   W.-J. Zhi, L.-F. Wang and X.-J. Hu, *Mil. Med. Res.,* 2017, **4**, 29.
3  K. Zhang, G.-H. Li, L.-M.Feng, N. Wang, J. Guo, K. Sun, K.-X. Yu, J.-B. Zeng, T. Li, Z. Guo and M. Wang, *J. Mater. Chem. C,* 2017, **5**, 9359.
4  Y. Zhan, M. Oliviero, J. Wang, A. Sorrentino, G. Buonocore, L. Sorrentino, M. Lavorgna, H. Xia and S. Iannace, *Nanoscale,* 2019, **11**, 1011.
5  H. Abbasi, M. Antunes and J. I. Velasco, *Prog. Mater. Sci.*, 2019, **103**, 319.
6  R. Pandey, S.Tekumalla and M. Gupta, *Compos. Part B-Eng.*, 2019, **163**, 150.
7  J. Jung, H. Lee, I. Ha, H. Cho, K. K. Kim, J. Kwon, P. Won, S. Hongand S. H. Ko, *ACS Appl. Mater. Interfaces,* 2017, **9**, 44609.
8  J. Huang, Z. Chen, F. Zhou, H. Wang, Y. Yuan, W. Chen, M. Gao and Y. Zhan, J. *Solid State Electrochem.,* 2017, **21**, 1559.
9  Y. Wang, W. Zhang, X. Wu, C. Luo, Q. Wang, J. Li and L. Hu, *Synthetic Met.,* 2017, **228**, 18.
10 J. Li, G. Zhang, H. Zhang, X. Fan, L. Zhou, Z. Shang and X. Shi, *Appl. Surf. Sci.,* 2018, **428**, 7.
11 M. H. Al-Saleh and U. Sundararaj, *Carbon,* 2009, **47**, 1738.
12 A. Gupta and V. Choudhary, *Compos. Sci. Technol.,* 2011, **71**, 1563.
13 L. Xu, X.-P. Zhang, C.-H. Cui, P.-G. Ren, D.-X. Yan and Z.-M. Li, *Ind. Eng. Chem. Res.,* 2019, **58**, 4454.
14 T. Hasan, Z. Sun, F. Wang, F. Bonaccorso, P. H. Tan, A. G. Rozhin and A. C. Ferrari, *Adv. Mater.,* 2009, **21**, 3874.
15 P. Cataldi, F. Bonaccorso, A. E. Del Rio Castillo, V. Pellegrini, Z. Jiang, L. Liu, N. Boccardo, M. Canepa, R. Cingolani, A. Athanassiou and I. S. Bayer, *Adv. Electron. Mater.,* 2016**, 2**, 1600245
16 P. Cataldi, I. S. Bayer, G. Nanni, A. Athanassiou, F. Bonaccorso, V. Pellegrini, A. E. Del Rio Castillo, F. Ricciardella, S. Artyukhin, M.-A. Tronche, Y. Gogotsi and R. Cingolani, *Carbon,* 2016, **109**, 331.
17 Y. Li, B. Shen, D. Yi, L. Zhang, W. Zhai, X. Wei and W. Zheng, *Compos. Sci. Technol.,* 2017, **138**, 209.
18 M. Cao, C. Han, X. Wang, M. Zhang, Y. Zhang, J. Shu, H. Yang, X. Fang and J. Yuan, *J. Mater. Chem. C,* 2018, **6**, 4586.
19 E. Lago,  P. S. Toth,  G. Pugliese,  V. Pellegrinia and F. Bonaccorso, *RSC Adv.*, 2016,**6**, 97931
20 Y. Zhan, J. Wang, K. Zhang, Y. Li, Y. Meng, N. Yan, W. Wei, F. Peng and H. Xia, *Chem. Eng. J.,* 2018, **344**, 184.
21 F. Shahzad, M. Alhabeb, C. B. Hatter, B. Anasori, S. M. Hong, C. M. Koo and Y. Gogotsi, *Science,* 2016, **353**, 1137.
22 Z. Huang, S. Wang, S. Kota, Q. Pan, M. W. Barsoum and C. Y. Li, *Polymer,* 2016, **102**, 119.
23 Y. Tong, M. He, Y. Zhou, X. Zhong, L. Fan, T. Huang, Q. Liao and Y. Wang, *Appl. Surf. Sci.,* 2018, **434**, 283.
24 Y. Jiang, X. Xie, Y. Chen, Y. Liu, R. Yang and G. Sui, *J. Mater. Chem. C,* 2018**, 6**, 8679.
25 W. Cao, F. Chen, Y. Zhu, Y. Zhang, Y. Jiang, M. Ma and F. Chen, *ACS Nano,* 2018, **12**, 4583.

26  R. Sun, H. Zhang, J. Liu, X. Xie, R. Yang, Y. Li, S. Hong and Z. -Z. Yu, *Adv. Funct. Mater.,* 2017, **27**, 1702807.
27  L. Wang, H. Qiu, C. Liang, P. Song, Y. Han, Y. Han, J. Gu, J. Kong, D. Pan and Z. Guo, *Carbon,* 2019, **141**, 506.
28  C. Liang, H. Qi, Y. Han, H. Gu, P. Song, L. Wang, J. Kong, D. Cao and J. Gu, *J. Mater. Chem. C,* 2019, **7**, 2725.
29  Q. Weng, X. Wang, X. Wang, Y. Bando and D. Golberg, *Chem. Soc. Rev.,* 2016, **45**, 3989.
30  G. Cassabois, P. Valvin and B. Gil, *Nat. Photonics,* 2016, **10**, 262.
31  X. Zhang, X. Zhang, M. Yang, S. Yang, H. Wu, S. Guo and Y. Wang, *Compos. Sci. Technol.,* 2016, **136**, 104.
32  X. Zhang, J. Zhang, C. Li, J. Wang, L. Xia, F. Xu, X. Zhang, H. Wu and S. Guo, *Chem. Eng. J.,* 2017, **328**, 609.
33  C.-P. Feng, S.-S. Wan, W.-C. Wu, L. Bai, R.-Y. Bao, Z.-Y. Liu, M.-B. Yang, J. Chen and W. Yang, *Compos. Sci. Technol.,* 2018, **167**, 456.
34  M. N. Gueye, A. Carella, R. Demadrille and J.-P. Simonato *ACS Appl. Mater. Interfaces,* 2017, **9**, 27250.
35  Y. Liu, M. Lu, K. Wu, S. Yao, X. Du, G. Chen, Q. Zhang, L. Liang and M. Lu, *Compos. Sci. Technol.,* 2019, **174**, 1.
36  Y. Zhan, Y. Meng and Y. Li, *Mater. Lett.,* 2017, **192**, 115.
37  P. Zhang, X. Ding, Y. Wang, Y. Gong, K. Zheng, L. Chen, X. Tian and X. Zhang, *Compos. Part A- Appl. S.,* 2019, **117**, 56.
38  H. Chen, V. V. Ginzburg, Yang, Y. Yang, W. Liu, Y. Huang, L. Du and B. Chen, *Prog J.. Polym. Sci.,* 2016, **59**, 41.
39  Y. Zhan, M. Lavorgna, G. Buonocore and H. Xia, *J. Mater. Chem.,* 2012, **22**, 10464.
40  M. Salzano de Luna, Y. Wang, T. Zhai, L. Verdolotti, G. G. Buonocore, M. Lavorgna and H. Xia, *Prog. Polym. Sci.,* 2019, **89**, 213.
41  L.-C. Jia, D.-X. Yan, Y. Yang, D. Zhou, C.-H. Cui, E. Bianco, J. Lou, R. Vajtai, B. Li, P. M. Ajayan and Z.-M. Li, *Adv. Mater. Technol.,* 2017, **2**, 1700078.
42  F. Sharif, M. Arjmand, A.A. Moud, U. Sundararaj and E.P.L. Roberts, *ACS Appl. Mater. Interfaces*, 2017, **9**, 14171.
43  X. Wang and P. Wu, *ACS Appl. Mater. Interfaces,* 2017, **9**, 19934.
44  C. Gao, H. Lu, H. Ni, and J. Chen, *J. Polym. Res.,* 2018, **25**, 6.
45  C. Feng, H. Ni, J. Chen, and W. Yang, *ACS Appl. Mater. Interfaces*, 2016, **8**, 19732.
46  C. Bottier, Ad. *Bot. Res.*, 2020, **93**, 201
47  A. E. Del Rio Castillo, V. Pellegrini, A. Ansaldo, F. Ricciardella, H. Sun, L. Marasco, J. Buha, Z. Dang, L. Gagliani, E. Lago, N. Curreli, S. Gentiluomo, F. Palazon, M. Prato, R. Oropesa-Nuñez, P. S. Toth, E. Mantero, M. Crugliano, A. Gamucci, A. Tomadin, M. Polini and F. Bonaccorso, *Mater. Horiz.,* 2018, **5**, 890.
48  A. E. Del Rio Castillo, A. Ansaldo, V. Pellegrini and F. Bonaccorso, International Patent no. WO2017089987A1, Nov. 2015.
49  A. E. Del Río Castillo, C. D. Reyes-Vazquez, L. E. Rojas-Martinez, S. B. Thorat, M. Serri, A. L. Martinez-Hernandez C. Velasco-Santo, V. Pellegrini, F. Bonaccorso. *FlatChem,* 2019, **18**, 100131
50  S. Kwon, R. Ma, Kim, H.R. Choi and S. F. Baik, *Carbon*, 2014, **68**, 118.
51  K. Zhang, H. O. Yu, Y. D. Shi, Y. F. Chen, J. B. Zeng, J. Guo, B. Wang, Z. Guo and M. Wang, *J. Mater. Chem. C,* 2017, **5**, 2807.
52  S. Berber, Y.-K. Kwon and D. Tomanek, *Phys. Rev. Lett.,* 2000, **84**, 4613.
53  C. Yuan, B. Duan, L. Li, B. Xie, M.Huang and X. Luo, *ACS Appl. Mater. Interfaces,* 2015, **7**, 13000.
54  Y. Zhan, J. Wu, H. Xia, N. Yan, G. Fei and G. Yuan, *Macromol. Mater. Eng.,* 2011, **296**, 590.
55  A. A. Balandin, S. Ghosh, W. Bao, I. Calizo, D. Teweldebrhan, F. Miao and C. N. Lau, *Nano Lett.,* 2008, **8**, 902.
56  G.-D. Zhan, J. D. Kuntz, A. K. Mukherjee, P.Zhu and K. Koumoto, *Scripta Mater.,* 2006, **54**, 77.
57  H. Ishida and S. Rimdusit, *J. Therm. Anal. Calorim.,* 1999, **58**, 497.
58  N. Burger, A. Laachachi, M. Ferriol, M. Lutz, V. Toniazzo, and D. Ruch, *Prog. Polym. Sci.,* 2016, **61**, 1.
59  F. F. Komarov, P. Zukowski, R. M. Kryvasheyeu, E. Munoz, T. N. Koltunowicz, V. N. Rodionova and A. K. Togambaeva, *Phys. Status Solidi A*, 2015, **212**, 425.
60  S. Sankaran, K. Deshmukh, M. Basheer Ahamed, K. K. Sadasivuni, M. Faisal and S. K. Khadheer Pasha, *Polym. Plast. Technol.,* 2019, **58**, 1191.

61 S. Mondal, S. Ganguly, M. Rahaman, A. Aldalbahi, T. K. Chaki, D. Khastgir and N. Ch. Das, *Phys. Chem. Chem. Phys.,* 2016, **18**, 24591.

62 L. Zhang, M. Baima and T. L. Andrew, *ACS Appl. Mater. Interfaces,* 2017, **9**, 32299.

63 Q.-W. Wang, H.-B. Zhang, J. Liu, S. Zhao, X. Xie, L. Liu, R. Yang, N. Koratkar and Z.-Z. Yu, *Adv. Funct. Mater.,* 2019, **29**, 1806819.

64 Y. Guo, C. Dun, J. Xu, J. Mu, P. Li, L. Gu, C. Hou, C. A. Hewitt, Q. Zhang, Y. Li, D. L. Carroll and H. Wang, *Small,* 2017, **13**, 1702645.

65 J. An and Y.G. Jeong, *Eur. Polym. J.*, 2013, **49**, 1322.

66 Y.G. Jeong and J. An, *Compos. Part A*, 2014, **56**, 1.

67 J. Yan and Y.G. Jeong, *Compos. Sci. Technol*., 2015, 106, 134.